\documentstyle[12pt]{article}
\advance \topmargin by -\headheight
\advance \topmargin by -\headsep
     
\evensidemargin \oddsidemargin
\def\rf#1{(\ref{eq:#1})}
\def\lab#1{\label{eq:#1}}
\def\nonu{\nonumber}
\def\br{\begin{eqnarray}}
\def\er{\end{eqnarray}}
\def\be{\begin{equation}}
\def\ee{\end{equation}}
\def\eq{\!\!\!\! &=& \!\!\!\! }

\def\lb{\lbrack}
\def\rb{\rbrack}

\def\llb{\left\lbrack}
\def\rrb{\right\rbrack}

\def\({\left(}
\def\){\right)}
\def\v{\vert}                     

\def\bc{\begin{center}}
\def\ec{\end{center}}

\def\Tr{\mathop{\rm Tr}}                  
\newcommand\partder[2]{{{\partial {#1}}\over{\partial {#2}}}}

\newcommand\sbr[2]{\left\lbrack\,{#1}\, ,\,{#2}\,\right\rbrack} 
\newcommand\pbr[2]{\{\,{#1}\, ,\,{#2}\,\}}       

\def\d{\delta}

\def\h{{1\over 2}}
\def\l{\lambda}

\def\P{\Phi}
\def\pa{\partial}

\def\pr{\prime}

\def\th{\theta}

\newcommand\tmat[9]{\left(\begin{array}{ccc}  
{#1} & {#2} & {#3} \\
{#4} & {#5} & {#6} \\
{#7} & {#8} & {#9} \end{array} \right)}

\newcommand\tcol[3]{\left(\begin{array}{c}  
{#1} \\ {#2} \\
{#3} \end{array} \right)}
\def\cA{{\cal A}}

\def\cH{{\cal H}}

\def\cK{{\cal K}}

\def\cM{{\cal M}}

\def\cR{{\cal R}}

\def\lie{{\cal G}}

\def\dt{{D_{\th}}}
\def\kere{\mbox{\rm Ker (ad$_E$)}}
\def\ime{\mbox{\rm Im (ad$_E$)}}

\newcommand{\ct}[1]{\cite{#1}}
\newcommand{\bi}[1]{\bibitem{#1}}
\newcommand\CMP[3]{{\sl Commun. Math. Phys.} {\bf #1} (#2) #3}

\newcommand\PLA[3]{{\sl Phys. Lett.} {\bf #1A} (#2) #3}
\newcommand\PLB[3]{{\sl Phys. Lett.} {\bf #1B} (#2) #3}
\newcommand\JMP[3]{{\sl J. Math. Phys.} {\bf #1} (#2) #3}

\newcommand\FAaIA[3]{{\sl Functional Analysis and Its Application} {\bf #1}
(#2) #3}

\newcommand\IJMPA[3]{{\sl Int. J. Mod. Phys.} {\bf A#1} (#2) #3}

\newcommand\JPA[3]{{\sl J. Physics} {\bf A#1} (#2) #3}
\newcommand\JSM[3]{{\sl J. Soviet Math.} {\bf #1} (#2) #3}
\newcommand\MPLA[3]{{\sl Mod. Phys. Lett.} {\bf A#1} (#2) #3}

\begin{document}
\begin{titlepage}
\vspace*{-1cm}
\noindent
April, 1997  \hfill{UICHEP-TH/97-5} \\
${}$ \hfill{hep-th/9704119}

\vspace {1.8cm}

\begin{center}
{\Large {\bf Zero Curvature Formalism for Supersymmetric 
Integrable Hierarchies in Superspace}}
\end{center}
\vskip .3in
\begin{center}
{  H. Aratyn\footnotemark
\footnotetext{Work supported in part by U.S. Department of Energy,
contract DE-FG02-84ER40173}${}^{,\,\rm a}$, A. Das\footnotemark
\footnotetext{Work supported in part by U.S. Department of Energy,
contract DE-FG02-91ER40685}${}^{,\,\rm b}$ and 
C. Rasinariu${}^{\,\rm a}$}

\par \vskip .1in \noindent
${}^{\rm (a)}$Department of Physics \\
University of Illinois at Chicago\\
845 W. Taylor St.\\
Chicago, IL 60607-7059
\par \vskip .1in \noindent

${}^{\rm (b)}$Department of Physics and Astronomy\\
University of Rochester\\
Rochester, N.Y. 14627
\end{center}

\vspace{2cm}

\begin{abstract}
We generalize the Drinfeld-Sokolov formalism of bosonic integrable hierarchies
to superspace, in a way which systematically leads to the zero curvature 
formulation for the supersymmetric integrable systems starting from the 
Lax equation in superspace. 
We use the method of symmetric space as well as the non-Abelian 
gauge technique to obtain the supersymmetric integrable hierarchies of the 
AKNS type from the  zero curvature condition in superspace with the graded 
algebras, $sl(n+1,n)$,  providing the Hermitian symmetric space structure.
\end{abstract}
\end{titlepage}

\section{Introduction}
Integrable systems have played an increasingly important role in 
the study of string theories. 
In fact, in this connection, it is the supersymmetric
integrable systems which become naturally more relevant. 
Although many properties of the bosonic integrable systems are quite well 
understood, the same is not true for supersymmetric systems. 
For example, it is well known that the bosonic integrable systems can be 
described naturally through Lax operators which are pseudo-differential 
operators \ct{GD}. 
It is also known through the works of Drinfeld and Sokolov (DS) \ct{D-S} 
that given a Lax description of a bosonic integrable system, 
one can systematically obtain from it a zero curvature formulation of
the same system where the matrices (gauge potentials) belong to the symmetry
group $sl(n)$ with $n$ related to the order of the pseudo-differential
operator.
The Drinfeld-Sokolov formalism connecting the scalar Lax formalism to
the matrix zero curvature formalism is quite important for it shows that one
can uniquely associate with every integrable nonlinear partial differential
equation, a unique affine Kac-Moody algebra \ct{D-S,wilson}. 
The zero curvature formulation depends crucially on the grading of 
the $sl(n)$ algebra. 
While the work of DS uses the principal grading of the underlying algebra, 
an alternate homogeneous grading leads to models of AKNS type \ct{FK83,AGZ}. 
An essential element in all these algebraic constructions of the integrable 
hierarchies is the semi-simple (diagonalizable) element $\Lambda$. 
For the AKNS models, $\Lambda = \lambda E$ is taken to be non-regular element 
while in the original work of DS, it is assumed to be regular. 
In the case of the AKNS models, a further simplification
occurs in that the construction of the integrable hierarchies can be carried
out in a straightforward manner using the symmetric space technique.

In contrast, while the supersymmetric integrable systems are well understood 
in the Lax formalism with scalar pseudo-differential operators in superspace
\ct{BK87}, most of the zero curvature formulations are given 
in components \ct{gurses}-\ct{dr3} (see however \ct{inami,toppan}). 
The underlying symmetry groups in the zero curvature formulation turn out to 
be the graded $sl$ or $osp$ groups as one would expect, but the actual zero 
curvature formulations are carried out by brute force. 
In other words, a satisfactory generalization of the DS formalism in the
superspace does not yet exist which would systematically associate with a
scalar Lax operator in superspace, a zero curvature condition and thereby
associate with a set of integrable supersymmetric nonlinear partial
differential equation a unique super Kac-Moody algebra.

In this letter, we propose such a generalization for the models of AKNS type.
Namely, we work with manifestly supersymmetric $1+1$ dimensional
(constrained) KP hierarchies which have superfield formulation with
$N$ (Extended) supersymmetry. 
In section 2, we first describe how one can systematically obtain a 
zero curvature formulation starting from a scalar Lax operator in
superspace. The underlying symmetry algebra naturally corresponds to 
the graded algebra, $sl(n+1,n)$. 
We, then, obtain the integrable hierarchies from the superspace 
zero curvature condition using the method of symmetric space and the
non-Abelian gauge technique. 
In section 3, we work out explicitly the case of
$N=1$ super KP hierarchy as an example of the general ideas of section 2. 
In section 4, we work out the $N=2$ example in detail.  

\section{Zero-curvature Equation and Gauge Techniques
for the Supersymmetric Constrained Lax Operator}
\subsection{Matrix Eigenvalue Problem for the Supersymmetric Constrained
Lax Operator}  
We first describe a method of associating a matrix eigenvalue equation
to the eigenvalue equation for the pseudo-differential Lax operator $L$
of the grade zero of the following form \ct{AR,ZP}
\br
L \psi_{BA} (t, \th_1,\ldots , \th_N ) \eq 
\(\pa + \sum_{a=1}^m \sum_{j=1}^N \P_a (t, \th_1,\ldots , \th_N ) 
\dt_j^{-1} \Psi_a (t, \th_1,\ldots , \th_N  ) \)
\psi_{BA} (t, \th_1,\ldots , \th_N ) \nonu \\
&= &\l \psi_{BA} (t, \th_1,\ldots , \th_N ) 
\lab{eigenl}
\er
where the superfields $\P_a (t, \th_1,\ldots , \th_N )$ and
$\Psi_a (t, \th_1,\ldots , \th_N  )$ are, respectively,
eigenfunctions and  adjoint  eigenfunctions
of $L$, satisfying
\be
\pa_n \P_a =\( L\)^n_+ \P_a~;\quad 
\pa_n \Psi_a = - \( L^*\)^n_+ \Psi_a~;\quad n=1,2,\ldots
\lab{eigen-f}
\ee
In the sequel we will choose the following grading
$\v \P_a \v = 0$ and $\v \Psi_a \v =1$.
Moreover, $\dt_j $ are covariant derivatives of the form:
$ \dt_j = \partder{}{\th_j} + \th_j \pa$, with $ \pa f = \pa f/ \pa x $.
They satisfy $\dt_j^2 = \pa$ for all $j =1, \ldots , N$.
Let $ S_a ( t, \th_1,\ldots , \th_N )$ 
be a function such that
\be
\pa S_a ( t, \th_1,\ldots , \th_N )
 =
 \Psi_a (t, \th_1,\ldots , \th_N  ) \psi_{BA} (t, \th_1,\ldots , \th_N  ) 
\lab{sfct}
\ee
As described in \ct{s-trick} (see also \ct{cheng}) $ S_a $ exists and is 
defined up to a constant.
In terms of the functions in \rf{sfct} the eigenvalue equation \rf{eigenl}
takes a form:
\br
\pa \psi_{BA} (t, \th_1,\ldots , \th_N ) &+&
\sum_{a=1}^m \sum_{j=1}^N \P_a (t, \th_1,\ldots , \th_N ) 
\dt_j S_a ( t, \th_1,\ldots , \th_N  ) \nonu \\
&=&\l \psi_{BA} (t, \th_1,\ldots , \th_N ) 
\lab{eigenla}
\er
We now associate to the eigenvalue problem a column vector
$\chi$:
\br
\chi^{T} (t, \th_1,\ldots , \th_N  )
&\equiv& \( \psi_{BA},  \dt_{1}  \psi_{BA}, \ldots,
\dt_{N}  \psi_{BA}, \ldots , \dt_{1} \dt_{2}\cdots \dt_{N} \psi_{BA}, \right. 
\lab{colvec} \\
& \dt_{1} S_1 &\!\!\! \!\!\left., \ldots, \dt_{1} S_m , 
\dt_{2} S_1 , \ldots, \dt_{2} S_m ,\ldots,
\dt_{1} \dt_{2}\cdots \dt_{N} S_m \)
\nonu
\er
where we have not explicitly shown all the ordered terms 
$\dt_{j_{1}} \cdots \dt_{j_{k}}\, , \, k \leq N $ such that 
$1 \leq j_{1} < j_{2} < \ldots < j_{k} \leq N$.
In this way $\chi$ contains $ 1+ m (2^N-1) + (2^N-1)= 2^N + m (2^N-1)$
elements and
we find a  one to one correspondence between the pseudo-differential 
Lax eigenvalue equation \rf{eigenl} and the first order matrix equation:
\be
\( \pa \cdot I - \l E + A \) \chi = 0
\lab{mateigen}
\ee
where $E= {\rm diag} \( \underbrace{1, \ldots , 1}_{2^N},
 \underbrace{0 , \ldots , 0}_{m (2^N-1)},\)$ and
$A$ is a $\(2^N + m (2^N-1)\) \times \(2^N + m (2^N-1)\)$
matrix which depends only on the superfields  appearing in $L$.
Since both matrices $E$ and $A$ have zero super-trace, 
we notice the appearance of the
$sl \( 2^{N-1} +m 2^{N-1}, m (2^{N-1} -1)+2^{N-1} \)$ algebraic structure.
Especially for the case $m=1$ we find the $sl (n+1,n)$ structure
with $n = 2^N -1$.

Let us furthermore describe the higher flows in terms of the matrix
equation:
\be
\(\partder{}{t_n}  + B_n \) \chi\, = \,0     \quad ; \quad n=2, 3, \ldots 
\lab{tnbn}
\ee
Assuming commutativity of flows $\sbr{\pa_x}{\partder{}{t_n}}=0$
we arrive at the usual zero-curvature (or Zakharov-Shabat) equation:
\be
\pa_n A - \pa B_n + \l \lb E \, , \, B_n \rb
-\lb A \, , \, B_n \rb = 0
\lab{zsa}
\ee
This, therefore, provides the generalization of the DS formalism to the
superspace.
\subsection{The Gauge Technique of Solving 
the Zero-Curvature equations}
To establish notation we will now present few facts about
the hermitian symmetric spaces.

Given is the algebra $\lie$ containing a semi-simple element
$E \in \lie$, which allows the decomposition
$\lie = {\rm Ker} ( {\rm ad}_E) \oplus {\rm Im } ( {\rm ad}_E)$ where 
``ad" represents the adjoint map.
Introduce, notation $\cK \equiv \kere$ and $\cM \equiv  \ime$.
We work here with $\lie / \cK$ being a Hermitian symmetric space, i.e.
satisfying 
\be
\lb \cK\, ,\, \cK \rb  \in \cK,\quad \lb \cM \, , \, \cK \rb \in \cM, \quad
\lb \cM \, , \, \cM \rb \in \cK,
\lab{A7}
\ee
as well as  
\be
( {\rm ad}_E)^2 {}_{\v_{\cM}} = I {}_{\v_{\cM}}.
\lab{ade2}
\ee
The well-known example of such algebras $\lie$ are $sl (n), n\geq 2$, 
which underlie the generalized AKNS models \ct{FK83,AGZ}.
In our discussions,  we will be interested in the example of the graded 
algebras $sl (n+1,n)$ with $n \geq 1$.
As we will see later, a specially relevant example is provided by $sl (2,1)$ 
with the (super-)traceless semi-simple element:
\be
E = \tmat{1}{0}{0}{0}{0}{0}{0}{0}{1}
\lab{els21}
\ee
In this case, the spaces $\ime$ and $\kere$ can be easily checked to have the
form
\be
\kere = \tmat{a_1}{0}{a_4}{0}{a_2}{0}{a_5}{0}{a_3}
\; ; \; 
\ime =  \tmat{0}{b_1}{0}{b_2}{0}{f_1}{0}{f_2}{0}
\lab{kmsl21}
\ee

We now turn our attention to the eigenvalue eq.
\rf{mateigen} and work with the general case of 
\be
A = A_{\cK} + A_{\cM} \in \ime + \kere = \lie
\lab{aimek}
\ee
We will make only one restriction on $A$, namely we will require
that the number of independent components of the (adjoint) eigenfunctions
in  $A$ does not exceed the algebraic dimensionality of $\ime$.
Or in other words we will require that the phase space is spanned by the
matrix elements of $A_{\cM}$ (namely, $b_1, b_2, f_1, f_2$ for the example in 
eq.\rf{kmsl21}) and, consequently, the matrix elements of $A_{\cK}$ depend 
functionally on those of $A_{\cM}$.

We now notice that the gauge potential in the direction of ${\cal K}$ can be
completely gauged away so that the eigenvalue eq.
\rf{mateigen} can be rewritten as
\be
\( \pa \cdot I  - \l E  + {\cA}\) {\bar \chi} = 0
\lab{mateigenb}
\ee
with
\br
{\cA} &\equiv& 
G^{-1} A G + G^{-1} \pa G = G^{-1} A_{\cM}  G
\lab{abarm}\\
{\bar \chi} &\equiv& G^{-1} \chi
\lab{chiabar}
\er  
where $G$ is determined from the condition:
\be
G^{-1} A_{\cK} G + G^{-1} \pa G =0  \quad ; \quad 
G = \exp \( -\int^x A_{\cK} dx\) 
\lab{g-def}
\ee
Because of condition \rf{g-def} it holds that
the matrix $\cA$ belongs to the image of ${\rm ad}_E$ \ct{AGZ}:
\be
\cA \in \ime.
\lab{ainime}
\ee
We turn our attention back to the zero-curvature eq.\rf{zsa}
derived from the eigenvalue problem \rf{mateigenb} and containing
$A$ substituted by $\cA$ as in \rf{ainime}. We note that the gauge
transformation was chosen to rotate the ${\cal K}$ component of $A$ from the
Lax equation, but, in general, this transformation would lead to both the
components in $B$ and, therefore,
we split the $B_n$ matrix into its $\kere$ and $\ime$ components
\be
B_n = B_n^{\cK} + B_n^{\cM}
\lab{bmsplit}
\ee
Accordingly, the zero-curvature eq.\rf{zsa} splits into
\br
\pa B_n^{\cK}+\lb \cA \, , \, B_n^{\cM} \rb & =& 0 \lab{kdir}\\
\pa_n \cA - \pa B_n^{\cM} + \l \lb E \, , \, B_n^{\cM}  \rb
-\lb \cA \, , \, B_n^{\cK} \rb & =& 0  \lab{mdir}
\er
along the $\kere$ and $\ime$ directions.
Equation \rf{kdir} can readily be integrated to yield
\be
B_n^{\cK} = - \pa^{-1}\(\lb \cA \, , \, B_n^{\cM} \rb\) - \l^n E
\lab{bnck}
\ee
where the last term is an \lq\lq integration constant'' chosen to ensure 
agreement with the limiting case $n=1$ for which $B_1 = \cA - \l E$.
Plugging eq.\rf{bnck} into eq.\rf{mdir} we get
\be
\pa_n \cA - \pa B_n^{\cM} + \l \lb E \, , \, B_n^{\cM}  \rb
+\llb \cA \, , \, \pa^{-1}\(\lb \cA \, , \, B_n^{\cM} \rb\) \rrb 
- \l^n \lb E \, , \, \cA \rb= 0
\lab{mdira}
\ee
We search for solutions of \rf{mdira} of the form
\be
B_{n}^{\cM} = \sum_{i=0}^{n-1} \l^{i} B^{\cM}_{n} (i)
\lab{1.8}
\ee
{}From \rf{mdira} we obtain for coefficients of $\l^n$:
\be
B^{\cM}_{n} (n-1) = \cA
\lab{bnm1}
\ee
after use of relation \rf{ade2}.
We also find from \rf{mdira} the recursion relation 
\be
B^{\cM}_{n} (i-1) = \cR B^{\cM}_{n} (i)
\lab{recbn}
\ee
for coefficients of expansion of $B^{\cM}_{n} $
with the recursion matrix given by
\be
\cR \equiv  ad_{E} \( \pa - ad_{\cA}\, \pa^{-1} ad_{\cA} \) 
\lab{recoper}
\ee
As a consequence we find
\be
{\rm ad}_E (\pa_n \cA ) = \cR ( B^{\cM}_{n} (0))= \cR^n (\cA) \quad ;\quad 
\pa_n \cA = ad_E \cR ad_E (\pa_{n-1} \cA)
\lab{recona}
\ee
Hence, in principle, once the recursion operator $\cR$ is determined, we can 
find all the evolution equations of the hierarchy.
Let us consider, for example, the first non-trivial case of $n=2$ for which we
find:
\be
{\rm ad}_E (\pa_2 \cA )= \cR^2 (\cA) = 
\pa^2_x \cA - \h {\rm ad}_E {\rm ad}_{\cA} {\rm ad}_{\cA} {\rm ad}_E (\cA)
\lab{pa2a}
\ee

{}From first of eqs.\rf{recona} we easily find
\be
{\rm ad}_E (\pa_{n-1} \cA ) = B^{\cM}_{n} (0) 
\to \pa B^{\cK}_{n} (0) = - {\rm ad}_{\cA} {\rm ad}_E (\pa_{n-1} \cA )
\lab{aux1}
\ee
where use was made of eq.\rf{bnck}.
Repeating arguments of \ct{AGZ} and \ct{wilson} we find that
\br
\pa_{n}\Tr \( \cA^{2} \)\eq
 2\Tr \( \lb E, \lb  E, \cA\rb \rb \pa_{n} \cA \)
= -2 \Tr \(E {\rm ad}_{\cA} {\rm ad}_E (\pa_{n} \cA ) \)
\nonu \\
\eq  2\pa \Tr \( E B^{\cK}_{n+1} (0) \)=  2 \pa \Tr \( E B_{n+1} (0) \)
\lab{hamden}
\er
Assuming now that $\cH_1 = \h \Tr \(\cA^{2}\)$ defines the first Hamiltonian 
density, it follows that $ \cH_{n} =   \Tr \( E B_{n+1} (0)\)$
satisfies the basic relation $ \pa \cH_n =  \pa_n \cH_1$.
Note also, that
$\Tr \( {\cA}^2 \) = \Tr \( A^2_{\cM}  \)$. This shows, in general, how we can
use the methods of symmetric space as well as the gauge technique to obtain the
hierarchy equations from the zero curvature condition. In the next two
sections, we will work out explicitly examples corresponding to the $N=1$ and
$N=2$ cases.

\section{Supersymmetric AKNS, Matrix Formulation and Gauging Technique}
The Lax operator of the supersymmetric AKNS model can be written as \ct{AR}:
\be
 L_{AKNS} = \pa + \P \dt^{-1} \Psi ~.
\lab{akns-lax}
\ee
According, to the discussion above introduce $S$ such that 
$ \pa_x S= \Psi \psi_{BA} $, and choose the
independent variables $\psi_{BA},\dt (S)$ and $\dt (\psi_{BA})$. 
Then eq. \rf{mateigen} is equivalent with
\be
\tmat{\pa - \l}{\P}{0}
         {-\dt( \Psi)}{\pa}{\Psi}
         {\P \Psi}{\dt \P}{\pa -\l}
\tcol{\psi_{BA}}{\dt (S)}{\dt (\psi_{BA})}=0~.
\lab{akns-mat}
\ee
If we use the matrix $E$ as in \rf{els21} and denote by
\be
A=\tmat{0}{\P}{0}
               {-\dt(\Psi)}{0}{\Psi}
               {\P \Psi}{\dt(\P)}{0}
\quad ; \quad
\chi=\tcol{\psi_{BA}}{\dt (S)}{\dt (\psi_{BA})}
\lab{sl21}
\ee
then eq. \rf{akns-mat} reads $\( \pa - \l E + A \) \chi = 0$
The matrices \rf{sl21} suggest an underlying sl(2,1)
algebraic structure.
We employ the gauge technique to bring the matrix $A$ from eq.\rf{sl21}
into the form of $\ime$ as in eq.\rf{kmsl21}.
To do it, we apply the gauge transformation from \rf{abarm} with
$G$ as in \rf{g-def}.
In this case it is given by
$G = \exp \(- \int^x \P \Psi \kappa_1 dx\)$ with
\be
\kappa_1 \, = \, \tmat{0}{0}{0}
                      {0}{0}{0}
                      {1}{0}{0}
\lab{t1}
\ee
After the gauge transformation performed as in eq.\rf{abarm} we obtain:
\be
\cA = \left(\begin{array}{ccr}
0 &  \Phi & 0 \\
-(\dt \Psi) + (\dt^{-2} \Phi \Psi) \Psi & 0 & \Psi \\
0 & (\dt \Phi) + (\dt^{-2} \Phi \Psi) \Phi & 0
\end{array}\right)
\lab{sl21g}
\ee
which can be plugged directly into eq.\rf{pa2a}
to obtain immediately the $n=2$  evolution equation of the hierarchy
for $\P, \Psi$:
\br
\pa_2 \P &=& \P^{\pr \pr} - 2 \( \P \Psi (\dt \P) - (\dt \Psi) \P^2 \)
\lab{pa2p}\\
\pa_2 \Psi &=& -\Psi^{\pr \pr} - 2 \, (\dt \Psi) \, \P \, \Psi 
\lab{pa2psi}
\er
These are, indeed, the equations of the hierarchy.
Note, that according to the discussion following eq.\rf{hamden}
the first two Hamiltonian densities are
\br
\cH_1 &=& \h \Tr \( A^2 \) =  \P {\bar \Psi} - {\bar \P} \Psi=
-  \( \P (\dt \Psi) + (\dt \P)\Psi \)
\lab{akns-h1} \\
\cH_2 &=& {\bar \Psi} \P^{\pr} - \P {\bar \Psi}^{\pr} + \Psi {\bar \P}^{\pr} 
- \Psi^{\pr} {\bar \P} \lab{akns-h2}
\er
where ${\bar \P} \equiv (\dt \Phi) + (\dt^{-2} \Phi \Psi) \Phi$
and ${\bar \Psi} \equiv -(\dt \Psi) + (\dt^{-2} \Phi \Psi) \Psi $.
The corresponding Poisson bracket structures for
the ``canonical'' variables $\P, {\bar \Psi}, \Psi , {\bar \P}$
are of the form:
\be
P_1 = \left(\begin{array}{cccc}
 0 & -1 &0&0 \\
1 &0&0&0 \\
 0&0&0 & -1\\
 0&0 & -1&0 \end{array} \right) 
\lab{akns-pb1}
\ee
and $P_2$ given by 
\be
\left(\begin{array}{cccc}
 -2 \P \pa^{-1} \P & -\pa - {\bar \P} \pa^{-1} \Psi +2 \P \pa^{-1} {\bar \Psi} 
 &   \P \pa^{-1} \Psi & - {\bar \P} \pa^{-1} \P - \P \pa^{-1} {\bar \P}    \\
-\pa +  \Psi\pa^{-1}{\bar \P}  +2 {\bar \Psi} \pa^{-1}  \P &
-2 {\bar \Psi} \pa^{-1} {\bar \Psi}& - {\bar \Psi} \pa^{-1} \Psi 
- \Psi \pa^{-1} {\bar \Psi}& {\bar \Psi} \pa^{-1} {\bar \P} \\
\Psi \pa^{-1} \P & - \Psi \pa^{-1} {\bar \Psi} -  {\bar \Psi} \pa^{-1} \Psi&0
 & \pa - {\bar \Psi} \pa^{-1} \P \\
- {\bar \P} \pa^{-1} \P -  \P \pa^{-1} {\bar \P} &
{\bar \P} \pa^{-1} {\bar \Psi} & - \pa +\P \pa^{-1}{\bar \Psi} &0 
\end{array} \right) 
\lab{akns-pb2}
\ee
where the matrix elements should be identified as
$P_{\{ij\}} \d (x-y) = \pbr{X_i(x)}{X_j(y)}$ for
$X = \P, {\bar \Psi}, \Psi , {\bar \P}$.
This bracket structures are consistent with expression $\cR = P_2 P_1^{-1}$
for the recursion matrix $\cR$ such that $\pa_n X = \cR \pa_{n-1} X$
They also lead to the correct equations of motion \rf{pa2p}-\rf{pa2psi}
via $ \pa_2 X = {\pbr{H_2}{X}}_2$ with $H_2 = \h \int dx \cH_2$.
Note, that it also holds that $ \pa_1 X = {\pbr{H_2}{X}}_1= {\pbr{H_1}{X}}_2$
with $H_1 =\int dx \cH_1$.
As explained in \ct{jacobi} presence of the non-local terms 
$\pa^{-1}_x\d (x-y)$ in \rf{akns-pb2} requires special care when verifying 
the Jacobi identities for the bracket $P_2$. 
We postpone dealing with this problem for a later publication.

\section{$N=2$ supersymmetric Lax}

In this section we illustrate our general formalism with an example possessing
$N=2$ supersymmetry.
Consider, namely the superspace described in terms of $(t,\th_1,\th_2)$, and 
the Lax operator:
\be
L = \pa + \P(t,\th_1,\th_2) \dt_1^{-1} \Psi(t,\th_1,\th_2)
  + \P(t,\th_1,\th_2) \dt_2^{-1} \Psi(t,\th_1,\th_2)~.
\lab{2-lax}
\ee
In order to obtain the hierarchy equations via the symmetric algebra formalism
we will associate to the eigenvalue equation
\be
   L \, \psi_{BA}(t,\th_1,\th_2) = \l \, \psi_{BA}(t,\th_1,\th_2)
\lab{baa}
\ee
a matrix eigenvalue equation $\( \pa \cdot I - \l E + A \) \chi = 0$
with
\be
E = \rm{diag} \( 1,1,0,0,0,1,1 \)
\lab{e-ZP}
\ee
and
\be
\chi^T = \( \psi_{BA}, \dt_1 \dt_2 \psi_{BA}, \dt_1 S, \dt_2 S,
\dt_1 \dt_2 S, \dt_1 \psi_{BA},  \dt_2 \psi_{BA}\)
\lab{chi-n2}
\ee
We find 
\be
A = \left(\begin{array}{ccccccc}  
0         & 0     & \P    & \P    & 0          & 0          & 0 \\
(D_1-D_{2}) \P \Psi & 0     & D_{12}\P & D_{12}\P & -(D_1+D_2) \P& 
-\P\Psi & \P\Psi\\
-D_1 \Psi  & 0     & 0       & 0       & 0          & \Psi  & 0 \\
-D_2 \psi  & 0     & 0       & 0       & 0          & 0     & \Psi \\
-D_{12}\Psi&-\Psi & 0      & 0       & 0          & -D_2 \Psi & D_1\Psi  \\
\P \Psi & 0   & D_1 \P   & D_1 \P    & \P     & 0          & 0 \\
\P \Psi & 0   & D_2 \P   & D_2 \P    & -\P     & 0 & 0 \end{array} \right)
\lab{n=2}
\ee
where for brevity we used $D_i = \dt_i \, , \, i=1,2$ and
$D_{12}= \dt_1 \dt_2$.

Applying the gauge technique within the setting of $sl(4,3)$ algebra
sends $A \to \cA \in \ime$ and 
results in the $N=2$ AKNS hierarchy evolution equations:
\br
\pa_2 \P &=& \P^{\pr \pr} - 2 \( \P \Psi (\dt_1 +\dt_2) \P 
-  \P^2 (\dt_1 +\dt_2)\Psi\)
\lab{pa2pn2}\\
\pa_2 \Psi &=& -\Psi^{\pr \pr} - 2  \, \P \, \Psi \, (\dt_1 +\dt_2) \Psi
\lab{pa2psin2}
\er

\section{Conclusion}
In this letter, we have generalized the Drinfeld-Sokolov formalism to
superspace which allows us to systematically construct a zero curvature
condition associated with a supersymmetric integrable system described by a
scalar Lax operator in superspace. We have shown, in general, how the AKNS type
of models in superspace can be obtained from the zero curvature condition
through the method of symmetric space and gauge technique. We have illustrated
the formalism by explicitly working out two examples with $N=1$ and $N=2$
supersymmetry.

\end{document}